\documentstyle[11pt,epsf]{article}

\newcommand{\beq}{\begin{equation}} \newcommand{\eeq}{\end{equation}}
\newcommand{\bea}{\begin{eqnarray}} \newcommand{\eea}{\end{eqnarray}}

  \newcommand
{\Romannumeral}[1]{\uppercase\expandafter{\romannumeral#1}}

\newcommand{\be}{\begin{enumerate}} \newcommand{\ee}{\end{enumerate}}
\newcommand{\bi}{\begin{itemize}} \newcommand{\ei}{\end{itemize}}
\newcommand{\ba}{\begin{array}} \newcommand{\ea}{\end{array}}
\newcommand{\bc}{\begin{center}} \newcommand{\ec}{\end{center}}
\newcommand{\bt}{\begin{tabular}} \newcommand{\et}{\end{tabular}}

\def\lsim{\mathrel{\rlap{\lower4pt\hbox{\hskip1pt$\sim$}}
    \raise1pt\hbox{$<$}}}           
\def\gsim{\mathrel{\rlap{\lower4pt\hbox{\hskip1pt$\sim$}}
    \raise1pt\hbox{$>$}}}           

\newcommand{\Tr}{\mathop{\rm Tr}}           
\newcommand{\tr}{\mathop{\rm tr}}           
\newcommand{\half}{\textstyle {1\over2} \displaystyle}    
\newcommand{\Dslash}{{\hbox{D}\kern-0.6em\raise0.15ex\hbox{/}}} 

\renewcommand{\et}{\eta}

\begin{document}

\setlength{\oddsidemargin}{0cm} \setlength{\baselineskip}{7mm}

\input epsf

\begin{normalsize}\begin{flushright}

May 2007 \\

\end{flushright}\end{normalsize}

\begin{center}
  
\vspace{5pt}

{\Large \bf Gravitational Wilson Loop and Large Scale Curvature}

\vspace{40pt}

{\sl H. W. Hamber}
$^{}$\footnote{On leave from the
Department of Physics, University of California, Irvine Ca 92717, USA.}
and 
{\sl R. M. Williams}
$^{}$\footnote{Permanent address:
Department of Applied Mathematics and Theoretical Physics,
Wilberforce Road, Cambridge CB3 0WA, United Kingdom.} \\

\vspace{30pt}

Max Planck Institute for Gravitational Physics \\
(Albert Einstein Institute) \\
D-14476 Potsdam \\
Germany \\

\vspace{20pt}

\end{center}

\begin{center} {\bf ABSTRACT } \end{center}

\noindent

In a quantum theory of gravity the gravitational Wilson loop, 
defined as a suitable
quantum average of a parallel transport operator around a large 
near-planar loop,
provides important information about the large-scale curvature properties of
the geometry.
Here we shows that such properties can be systematically computed
in the strong coupling
limit of lattice regularized quantum gravity, by performing local
averages over loop bivectors, and over lattice
rotations, using an assumed near-uniform measure in group space.
We then relate the resulting quantum averages to an expected semi-classical
form valid for macroscopic observers, which leads to an
identification of the gravitational correlation length appearing in the
Wilson loop with an observed large-scale curvature.
Our results suggest that strongly coupled gravity leads to
a positively curved (De Sitter-like) quantum ground state, 
implying a positive effective cosmological constant at large distances.





\vfill

\pagestyle{empty}

\newpage

\pagestyle{plain}

\section{Introduction}

\vskip 20pt

An important question for any theory of quantum gravity is what gravitational
observables should look like [1], i.e. which expectation values of operators 
(or ratios thereof) have meaning and physical interpretation in the context of 
a manifestly covariant formulation, specifically in a situation where metric 
fluctuations are not necessarily bounded. Such averages naturally include 
expectation values of the (integrated) scalar curvature and other related 
quantities (involving for example curvature-squared terms), as well as 
correlations of operators at fixed geodesic distance, sometimes referred to 
as bi-local operators. 
Another set of 
physical observables corresponds to the gravitational analog of the Wilson 
loop [2], providing information about the parallel transport of vectors, and 
therefore on the effective curvature, around large, near-planar loops, and 
the correlation between particle world-lines [3,4] (providing information about the 
static gravitational potential). In this paper we will concentrate on defining 
and exploring physical properties of the gravitational Wilson loop [5].

Before embarking on the gravitational case, let us recall more generally the
well-known fact that many low energy physical properties in gauge theories 
cannot be computed reliably in weak coupling perturbation theory.
Thus, for example, in non-Abelian gauge theories a confining potential 
for static sources placed in the fundamental representation 
is found at sufficiently strong coupling,
by examining the behavior of the Wilson loop [2], 
defined for a large closed planar loop $C$ as
\beq
W( C ) \, = \, 
< \tr {\cal P} \exp \Bigl \{ i g \oint_{C} A_{\mu} (x) dx^{\mu} 
\Bigr \} > \;\; ,
\label{eq:wloop_sun}
\eeq
with $A_\mu \equiv t_a A_\mu^a $ and the $t_a$'s the group
generators of $SU(N)$ in the fundamental representation. 
Specifically, in the pure gauge theory at strong coupling
the leading contribution to the Wilson loop can be shown to follow
an area law for sufficiently large loops
\beq
< W( C ) > 
\; \mathrel{\mathop\sim_{ A \, \rightarrow \, \infty  }} \;
\exp ( - A(C) / \xi^2 ) \;\; , 
\label{eq:wloop_sun1}
\eeq
where $A(C)$ is the minimal area spanned by the planar loop $C$ [6,7].
The quantity $\xi $ is the gauge field correlation length, defined 
for example from the exponential decay of the Euclidean correlation
function of two infinitesimal loops separated by a distance $|x|$,
\beq
G_{\Box} ( x ) \, = \, 
< \tr {\cal P} \exp \Bigl \{ i g \oint_{C_\epsilon'} A_{\mu} (x') dx'^{\mu} 
\Bigr \} (x)
\;
\tr {\cal P} \exp \Bigl \{ i g \oint_{C_\epsilon''} A_{\mu} (x'') dx''^{\mu} 
\Bigr \} (0)
>_c \;\; .
\label{eq:box_sun}
\eeq
Here the $C_\epsilon$'s are two infinitesimal loops centered around $x$ and $0$
respectively, suitably defined on the lattice as elementary square loops, 
and for which one has at sufficiently large separations
\beq
G_{\Box} ( x ) 
\; \mathrel{\mathop\sim_{ |x|  \, \rightarrow \, \infty  }} \;
\exp ( - |x| / \xi ) \;\; .
\label{eq:box_sun1}
\eeq
The inverse of the correlation length $\xi$ is known to correspond to the lowest
mass excitation in the gauge theory, the scalar glueball.

Not only will we adapt this definition to the gravitational case, but more 
specifically to the case of discrete gravity, in the context of the 
discretization scheme known as Regge calculus [8]. It will turn out 
that it is most easily achieved by using a slight variant of Regge calculus, 
in which the action coincides with the usual Regge action in the near-flat 
limit. In Section 2, we shall describe the lattice notion of parallel transport, 
and how areas are defined on the dual lattice. Then in 
Section 3, the gravitational Wilson loop will be defined, and it will be shown 
why, in the discrete case, we modify the action. In section 4, sample 
calculations will be performed and the behavior of large Wilson loops derived. 
Much of what is done will be in close parallel with the procedure in lattice 
gauge theories (as will be immediately obvious to those familiar with that area). 
We close with a discussion of the interpretation in semiclassical terms
of our main result for a large Wilson loop.
We will argue there that our results imply that for strong coupling
(large bare Newton's constant $G$) the behavior of the Wilson
loop is consistent with a positive vacuum curvature, and
therefore with (Euclidean) De Sitter space.

\vskip 40pt

\section{Rotations, parallel transport and Voronoi loops}

\vskip 20pt

In lattice gravity, space-time is built up from flat simplices, with curvature 
restricted to the subspaces of codimension 2. In four dimensions, this means 
that the hinges, where the curvature lies, are triangles. The contribution to 
the action of a hinge is the product of its area, $A_h$,
with the deficit angle, $\delta_h$, there. 
This is defined to be $2\pi$ minus the sum of the dihedral angles at that 
hinge, in the simplices meeting there. We may also define a volume associated 
with each hinge, $V_h$ (see later in this section). Then the lattice action 
for pure four-dimensional Euclidean gravity with a cosmological constant and 
the usual Einstein scalar curvature term is
\beq 
I_{latt} \; = \;  \lambda_0 \, \sum_h V_h  \, - \, 
k \sum_h \delta_h  \, A_h  \;\; , 
\label{eq:ilatt} 
\eeq
with $k=1/(8 \pi G)$, and $V_h, \;  A_h$ and $\delta_h$ are all functions of the  
edge lengths, which are the basic variables in the theory, analogous to the 
metric in the continuum.
This action only couples edges which belong either to
the same simplex or to a set of neighboring simplices, and can therefore
be considered as {\it local}, just like the continuum action. It 
leads to the regularized lattice functional integral [9]
\beq 
Z_{latt} \; = \;  \int [ d \, l^2 ] \; \exp \left \{  
- \lambda_0 \sum_h V_h \, + \, k \sum_h \delta_h A_h \right \} \;\; ,
\label{eq:zlatt} 
\eeq
where, as customary, the lattice ultraviolet cutoff is set equal to one
(i.e. all length scales are measured in units of the lattice cutoff).
The lattice partition function $Z_{latt}$ should then be compared to the
continuum Euclidean Feynman path integral 
\beq
Z_{cont} \; = \; \int [ d \, g_{\mu\nu} ] \; \exp \left \{  
- \lambda_0 \, \int d x \, \sqrt g \, + \, 
{ 1 \over 16 \pi G } \int d x \sqrt g \, R \right \} \;\; .
\label{eq:zcont}
\eeq

In practice the lattice functional integral $Z_{latt}$ should be regarded 
as a regularized form of the continuum Euclidean Feynman path integral.
The latter will involve a functional measure over metrics
$g_{\mu\nu}(x)$, usually of the form
\beq
\int [d \, g_{\mu\nu} ] \,  \equiv \, 
\prod_x \, \left [ g(x) \right ]^{\sigma / 2} \, 
\prod_{ \mu \ge \nu } \, d g_{ \mu \nu } (x) \;\; ,
\label{eq:cont-meas}
\eeq
where $\sigma$ is a parameter, constrained by the
requirement of a non-singular measure to $\sigma \ge - (d+1)$.
For $\sigma= \half (d-4)(d+1)$ one has De Witt's measure,
while for $\sigma= - (d+1)$ one recovers the original Misner measure.
Here we will mostly be interested in the physical four-dimensional
case, for which $d=4$ and therefore $\sigma=0$ in the De Witt measure;
the specific form of the functional measure is not expected to play an
important role in the following, except that it will assumed to
be diffeomorphism invariant [5].

Furthermore, unless stated otherwise, it will convenient
to include the cosmological constant term in the measure as well,
since this contribution is ultralocal and
contains no derivatives of the metric, giving rise to
an effective strong coupling measure $d \mu (l^2)$,
\beq 
d \mu (l^2) \; \equiv \; [ d \, l^2 ] \, e^{- \lambda_0 \sum_h V_h }
\;\; .
\label{eq:mulatt} 
\eeq
This last expression represents a fairly non-trivial 
quantity, both in view of the relative complexity
of the expression for the volume of a simplex,
and because of the generalized triangle inequality constraints
already implicit in the definition of $[d\,l^2]$.

The main assumption used here regarding this effective strong
coupling measure will be the existence of a stable ground
state with a well-defined average lattice spacing, as implied
by direct numerical evaluations of the lattice integrals in four
dimensions, at least for sufficiently strong coupling [5,10].
In the following the lattice measure will therefore be assumed 
to be a suitable discretization of the continuum functional measure,
and therefore of the form 
\beq
\int [ d \, l^2 ] \; = \;
\int_0^\infty \; \prod_s \; \left ( V_d (s) \right )^{\sigma} \;
\prod_{ ij } \, dl_{ij}^2 \; \Theta [l_{ij}^2] \;\; .
\label{eq:latt-meas}
\eeq
with $\sigma$ again a real parameter, and $\Theta$ a function of the
squared edge lengths, ensuring the validity of the triangle
inequalities.

At strong coupling the measure and cosmological constant terms
form the dominant part of the functional integral, since the Einstein
part of the action is vanishingly small in this limit.
Yet, and in contrast to strongly coupled lattice Yang-Mills theories,
the functional integral is still non-trivial to compute analytically in
this limit, mainly due to the triangle inequality constraints.
Therefore, in order to be able to derive some analytical estimates for
correlation functions in the strong coupling limit, one needs still
to develop some set of approximation methods, which will discussed below.
These methods and their results can later be tested by numerical means,
for example by integrating directly over edges, through the explicit 
lattice measure over edges given above.

One approach that appears natural in the gravity context
follows along the lines of what is normally done in gauge
theories, namely an integration over compact group variables,
using the invariant measure over the gauge group [2].
It is of this method that we wish to take advantage here, as we
believe that it is well suited for gravity as well.
In order to apply such a technique to gravity one needs (i) to
formulate the lattice theory in such a way that group variables are separated and
therefore appear
explicitly; (ii) integrate over the group variables using an invariant
measure; and (iii) approximate the relevant 
correlation functions in such a way
that the group integration can be performed exactly, using for example
mean field methods for the parts that appear less tractable.
In such a program one is aided, as will be shown below, by the fact that
in the strong coupling limit one is expanding about a well defined
ground state, and that the measure and the interactions are
{\it local}, coupling only lattice variable (edges or rotations) 
which are a few lattice spacings apart, thus excluding the appearance
of long-range (power-like) correlations.

The downside of such methods is that one is no longer evaluating the 
functional integral for quantum gravity exactly, even in the strong
coupling limit; the upside is that one obtains a clear analytical
estimate, which later can be in principle systematically tested by 
numerical methods (which are exact).

In the gravity case the analogs of the gauge variables of Yang-Mills
theories are given by the connection, so it is natural therefore
to look for a first order formulation of gravity.
In a first order formalism one writes for the Einstein-Hilbert pure gravity
Lagrangian density
\beq
{\cal L} \; = \; { 1 \over 16 \pi G } \, 
\sqrt{g} \, g^{\mu\nu} \, R_{\mu\nu} \;\; .
\eeq
For any spacetime manifold with an affine connection one has for the Ricci tensor
\beq
R_{ \mu \nu }  =  g^{ \lambda \sigma }  R_{ \lambda \mu \sigma \nu } \;\; ,
\eeq
where 
\beq
R_{ \; \mu \nu \sigma }^\lambda \; = \;
\partial_\nu \Gamma_{ \mu \sigma }^\lambda 
- \partial_\sigma \Gamma_{ \mu \nu }^\lambda
+ \Gamma_{ \mu \sigma }^\eta  \Gamma _{ \nu \eta }^\lambda
- \Gamma_{ \mu \nu }^\eta  \Gamma_{ \sigma \eta }^\lambda \;\; .
\eeq
Variation of the pure gravitational action requires that
\beq
{ 1 \over 16 \pi G } \, \int d^4 x \; \delta [ 
\sqrt{g} \, g^{\mu\nu} \, R_{\mu\nu} ] \; = \; 0  \;\; .
\eeq
The variation of the first and second terms inside the square parentheses
are trivial, and the variation of $ R_{\mu\nu} $
can be simplified by virtue of the Palatini identity
\beq
\delta \, R_{ \; \mu \nu \sigma }^\lambda \; = \;
\delta \Gamma_{ \mu \sigma ; \nu }^\lambda
- \delta \Gamma_{ \mu \nu ; \sigma }^\lambda \;\; .
\eeq
After integrating by parts one can then show that the term involving
the variation of the connection $\Gamma$ implies
\beq
\partial_\lambda g_{\beta\gamma} \, - \, 
g_{\gamma\sigma} \Gamma^\sigma_{\beta\lambda} \, - \, 
g_{\beta\sigma} \Gamma^\sigma_{\gamma\lambda} \; = \; 0
\eeq
(normally just written as $ g_{\mu\nu ; \lambda } = 0 $),
and which can be inverted to give the usual relationship
between the connection $\Gamma$ and the metric $g$ in Riemannian geometry,
namely
\beq
\Gamma_{ \mu \nu }^\lambda = \half \; g^{ \lambda \sigma }
\Bigl ( \partial_\mu g_{ \nu \sigma } 
+ \partial_\nu g_{ \mu \sigma } 
- \partial_\sigma g_{ \mu \nu } \Bigr ) \;\; .
\eeq
Equating to zero the coefficients of $\delta g_{\mu\nu}$
gives instead the ten components of Einstein's field equations.
The (well-known) conclusion therefore is that in the quantum
theory one can safely consider functionally integrating
separately over the affine connection and the metric, treated
as independent variables, with the correct relationship between
metric and connection arising then as a consequence of the dynamics.
For the rest of this section we will follow a similar spirit, separating
out explicitly in the lattice action the degrees of freedom
corresponding to local rotations (the analogs of the $\Gamma$'s in the
continuum), which we will find to be most conveniently described
by orthogonal matrices ${\bf R}$, implying a choice of preferred
coordinate systems within the simplices.

The next step is a discussion of the properties of local rotation matrices in the
context of the lattice theory, and how these
relate to the lattice gravitational action.
Since in Regge calculus the interior of each simplex $s$ is assumed to be flat, 
one can assign to it a Lorentz frame $\Sigma (s)$.
Furthermore inside $s$ one can define a $d$-component vector
$\phi (s) = ( \phi_0 \dots \phi_{d-1} )$.
Under a Lorentz transformation of $\Sigma (s)$, described by the
$d \times d$ matrix $\Lambda (s)$ satisfying the usual relation
for Lorentz transformation matrices $\Lambda^{T} \, \eta \, \Lambda \; = \; \eta$,
with $\eta$ the flat metric, the vector $\phi (s)$ will rotate to
\beq
\phi ' (s) \; = \; \Lambda (s) \, \phi (s) \; . 
\eeq
The base edge vectors $e_i^{\mu} = l_{0i}^{\mu} (s)$ themselves
are of course an example of such a vector.

Next consider two $d$-simplices, individually labeled by $s$ and $s'$,
sharing a common face $f (s,s')$ of dimensionality $d-1$ [10].
It will be convenient to label the $d$ edges residing in the common face
$f$ by indices $i,j=1 \dots d$.
Within the first simplex $s$ one can then assign a Lorentz frame $ \Sigma (s)$,
and similarly within the second $s'$ one can assign the frame $\Sigma (s')$.
The $\half d (d-1) $ edge vectors on the common interface
$f(s,s')$ (corresponding physically to the same edges, viewed from
two different coordinate systems)
are expected to be related to each other by a Lorentz rotation ${\bf R}$,
\beq
l_{ij}^\mu (s') \; = \; R_{\;\;\nu}^{\mu} (s',s) \; l_{ij}^{\nu} (s) \;\; .
\eeq
Under individual Lorentz rotations in $s$ and $s'$ one has of course a
corresponding change in $ R$, namely
${R} \rightarrow \Lambda (s') \, { R} (s',s) \, \Lambda (s)$.
In the Euclidean $d$-dimensional case $ R$ is an orthogonal matrix, an 
element of the group $SO(d)$.

In the absence of torsion, one can use the matrix ${ R}(s',s)$ to describes
the parallel transport of any vector $\phi$ from simplex $s$ to a
neighboring simplex $s'$,
\beq
\phi^\mu (s') \; = \; R_{\; \; \nu}^{\mu} (s',s) \, \phi^{\nu} (s)
\eeq
${R}$ therefore describes a lattice version of the connection [11,12].
Indeed in the continuum such a rotation would be described by the matrix
\beq
R_{\;\;\nu}^{\mu} \; = \; \left ( e^{\Gamma \cdot dx} \right )_{\;\;\nu}^{\mu}
\eeq
with $\Gamma^{\lambda}_{\mu\nu}$ the affine connection.
The coordinate increment $dx$ is interpreted as joining
the center of $s$ to the center of $s'$, thereby intersecting
the face $f(s,s')$.
Note that it is possible to choose coordinates so that
$ { R} (s,s')$ is
the unit matrix for one pair of simplices, but it will not then be unity for
all other pairs.

One can consider a sequence of rotations along an arbitrary
path $P (s_1, \dots , s_{n+1})$ going through simplices
$s_1 \dots s_{n+1}$, whose combined rotation matrix is given by
\beq
{\bf R} (P) \; = \; {R} (s_{n+1}, s_n ) \cdots {R} (s_2, s_1 )
\eeq
and which describes the parallel transport of an arbitrary vector
from the interior of simplex $s_1$ to the interior of simplex $s_{n+1}$,
\beq
\phi^\mu (s_{n+1}) \; = \; R_{\; \; \nu}^{\mu} (P) \, \phi^{\nu} (s_1)
\eeq
 If the initial and final simplices $s_{n+1}$ and $s_1$ coincide,
one obtains a closed path $C (s_1, \dots , s_n)$, for which the
associated expectation value can be considered as the gravitational
analog of the Wilson loop.
Its combined rotation is given by
\beq
{\bf R} (C) \; = \;  R (s_1, s_n ) \cdots  R (s_2, s_1 )
\label{eq:loop-rot}
\eeq
Under Lorentz transformations within each simplex $s_i$ along
the path one has a pairwise cancellation of the $\Lambda (s_i)$ matrices
except at the endpoints, giving in the closed loop case
\beq
{\bf R} (C) \; \rightarrow \; \Lambda ( s_1 ) \, {\bf R} ( C )
\, \Lambda^{T} ( s_1 )
\eeq
Clearly the deviation of the matrix ${\bf R} (C)$ from unity is a measure of curvature.
Also, the trace $\tr {\bf R} (C)$ is independent of the choice
of Lorentz frames.

\begin{center}
\epsfxsize=10cm
\epsfbox{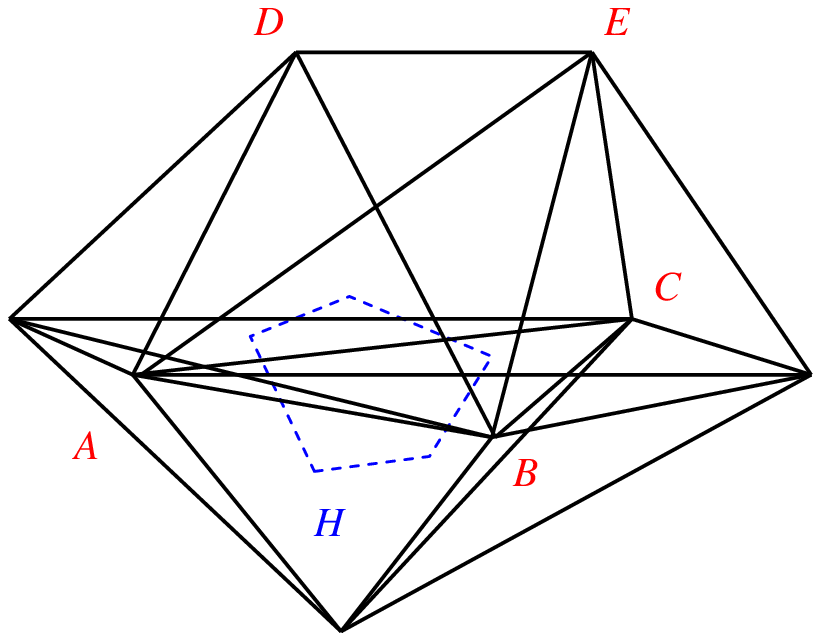}
\end{center}

\noindent{\small Figure 1.
Elementary polygonal path around a hinge
(triangle) in four dimensions. The hinge $ABC$, contained in the simplex
$ABCDE$, is encircled by the polygonal path $H$ connecting the
surrounding vertices which reside in the dual lattice.
One such vertex is contained within the simplex $ABCDE$.
\medskip}

Of particular interest is the elementary loop associated with
the smallest non-trivial, segmented parallel transport path one can
build on the lattice.
One such polygonal path in four dimensions is shown in
Figure 1.
In general consider a $(d-2)$-dimensional simplex (hinge) $h$, which
will be shared by a certain number $m$ of $d$-simplices,
sequentially labeled by $s_1 \dots s_m$, and whose common faces
$f(s_1,s_2) \dots f( s_{m-1}, s_m ) \; f(s_m, s_1 ) $ will also contain the hinge $h$.
In four dimensions several four-simplices will contain,
and therefore encircle, a given triangle (hinge).
In three dimensions the path will encircle an edge, while in two
dimensions it will encircle a site.
Thus for each hinge $h$ there is a unique elementary closed path $C_h$
for which one again can define the ordered product
\beq
{\bf R} (C_h) \; = \;  R (s_1, s_m ) \cdots  R (s_2, s_1 )
\label{eq:loop-rot1}
\eeq
The hinge $h$, being geometrically an object of dimension $(d-2)$, is naturally
represented by a tensor of rank $(d-2)$, referred to a coordinate system
in $h$: an edge vector $l_h^\mu$ in $d=3$, and an area bivector
$\half ( l_h^\mu l_h^{'\nu} - l_h^\nu l_h^{'\mu} ) $ in $d=4$ etc.
It will therefore be convenient
to define a hinge bivector $U$ in any dimension as
\beq
U_{\mu\nu} (h) \; = \; 
{\cal N} \, \epsilon_{\mu\nu \alpha_1 \alpha_{d-2}} \, 
l_{(1)}^{\alpha_1} \dots l_{(d-2)}^{\alpha_{d-2}}
\;\; ,
\label{eq:bivector-d}
\eeq
normalized, by the choice of the constant ${\cal N}$, in such a way that
$U_{\mu\nu} U^{\mu\nu} =2$.
In four dimensions
\beq
U_{\mu\nu} (h) \; = \; { 1 \over 2 A_h } \;
\epsilon_{\mu\nu\alpha\beta} \, l_1^{\alpha} \, l_2^{\beta}
\label{eq:bivector}
\eeq
where $l_1 (h)$ and $l_2 (h)$ two independent edge vectors
associated with the hinge $h$,
and $A_h$ the area of the hinge.

An important aspect related to the rotation of an
arbitrary vector, when parallel transported around a hinge $h$,
is the fact that, due to the hinge's intrinsic orientation,
only components of the vector in the plane perpendicular
to the hinge are affected.
Since the direction of the hinge $h$ is specified locally by
the bivector $U_{\mu\nu}$ of Eq.~(\ref{eq:bivector}),
one can write for the loop rotation matrix $\bf R$

\beq
R_{\;\;\nu}^{\mu} (C) \; = \;
\left ( e^{ \delta \, U } \right )_{\;\;\nu}^{\mu}
\label{eq:rot-hinge}
\eeq

\noindent where $C$ is now the small polygonal loop entangling the hinge $h$,
and $\delta$ the deficit angle at $h$.
One particularly noteworthy aspect of this last result is the fact that the area
of the loop $C$ does not enter in the expression for the
rotation matrix, only the deficit angle and the hinge direction.
Note that in the above expression for the rotation matrix ${\bf R}(C)$
both the deficit angle $\delta (C)$ giving the magnitude of the 
rotation, as well as the bivector $U(C)$ giving the direction of the rotation, 
are each rather complicated functions of the original edge 
lengths, with the latter also depending on a choice for the local
coordinate system.

At the  same time, in the continuum a vector $V$ carried around
an infinitesimal loop of area $A_C$ will change by
\beq
\Delta V^{\mu} \; = \; \, \half \, R^{\mu}_{ \;\; \nu \lambda \sigma }
\, A^{ \lambda \sigma } \, V^\nu
\eeq
where $A^{ \lambda \sigma }$ is an area bivector in the plane of $C$,
with squared magnitude $ A_{ \lambda \sigma } A^{ \lambda \sigma } = 2 A_C^2$.
Since the change in the vector $V$ is given by
$ \delta V^\alpha = ({\bf{R-1}})^{\alpha}_{\;\;\beta} \, V^\beta $
one is led to the identification
\beq
\half \; R^{\alpha}_{\;\;\beta\mu\nu} \, A^{\mu\nu} \; = \;
({\bf {R-1}})^{\alpha}_{\;\;\beta} \;\; .
\label{eq:riemrot}
\eeq
Consequently the above change in $V$ can equivalently be re-written in terms of the
infinitesimal rotation matrix
\beq
R_{\;\;\nu}^{\mu} (C) \; = \; 
\left ( e^{ \, \half \, R \cdot A } \right )_{\;\;\nu}^{\mu}
\eeq
(where the Riemann tensor appearing in the exponent on the r.h.s. should not
be confused with the rotation matrix $\bf R$ on the l.h.s.).

The area $A_C$ is most suitably defined by introducing the
notion of a {\it dual lattice},
i.e. a lattice constructed by assigning centers to the simplices,
with the polygonal curve $C$ connecting these centers sequentially,
and then assigning an area to the interior of this curve.
One possible way of assigning such centers is by introducing
perpendicular bisectors to the faces of a simplex, and locating
the vertices of the dual lattice at their common intersection,
a construction originally discussed in [13,14].
Another, and perhaps even simpler, possibility is to use a barycentric
subdivision. Then the volume element, $V_h$, is 
defined by first joining the vertices of the
polyhedron $C$, whose vertices lie in the dual lattice,
with the vertices of the hinge $h$, and then computing its volume.
We can then show that the polygonal area $A_C$ is given by 
$A_C (h) = d \; V_h / V^{(d-2)}_h $, where $V^{(d-2)}_h$
is the volume of the hinge (a triangle area in four dimensions).

\vskip 40pt

\section{Gravitational Wilson loop}

\label{sec:loop}

\vskip 20pt

We have seen that with each neighboring pair of simplices $s,s+1$ one 
can associate a
Lorentz transformation $ R^{\mu}_{\;\; \nu} (s,s+1)$, which describes
how a given vector $ V^\mu $ transforms between the local coordinate
systems in these two simplices, and
that the above transformation is directly related to the continuum
path-ordered ($P$) exponential of the integral of the local affine connection
$ \Gamma^{\lambda}_{\mu \nu}(x)$ via
\beq
R^\mu_{\;\; \nu} \; = \; \Bigl [ P \; e^{\int_
{{\bf path \atop between \; simplices}}
\Gamma_\lambda d x^\lambda} \Bigr ]^\mu_{\;\; \nu}  \;\; .
\eeq
with the connection having support only on the common interface between the two
simplices.
Also, for a closed elementary path 
$C_h$ encircling a hinge $h$ and passing through
each of the simplices that meet at that hinge one has
for the total rotation matrix ${\bf R} \equiv \prod_s  R_{s,s+1} $
associated with the given hinge 
\beq
\Bigl [ \prod_s  R_{s,s+1}   \Bigr ]^{\mu}_{\;\; \nu} \; = \;
\Bigl [ \, e^{\delta (h) U (h)} \Bigr ]^{\mu}_{\;\; \nu}  \;\; ,
\eeq
\noindent as in Eq.~(\ref{eq:rot-hinge}). This matrix describes the parallel 
transport of a vector round the loop.

More generally one might want to consider a near-planar, but non-infinitesimal,
closed loop $C$, as shown in Figure 2.
Along this closed loop the overall rotation matrix will still be given by 
\beq
R^{\mu}_{\;\; \nu} (C) \; = \;
\Bigl [ \prod_{s \, \subset C}  R_{s,s+1} \Bigr ]^{\mu}_{\;\; \nu} 
\eeq
In analogy with the infinitesimal loop case,
one would like to state that for the overall rotation matrix one has
\beq
R^{\mu}_{\;\; \nu} (C) \; \approx \; 
\Bigl [ \, e^{\delta (C) U (C))} \Bigr ]^{\mu}_{\;\; \nu}  \;\; ,
\label{eq:latt-wloop}
\eeq
where $U_{\mu\nu} (C)$ is now an area bivector perpendicular to the
loop, which will work only if the loop is close to planar so
that $U_{\mu\nu}$ can be taken to be approximately constant
along the path $C$. By a near-planar loop around the point $P$, we mean 
one that is constructed by drawing outgoing geodesics, on a plane through $P$.

\begin{center}
\epsfxsize=10cm
\epsfbox{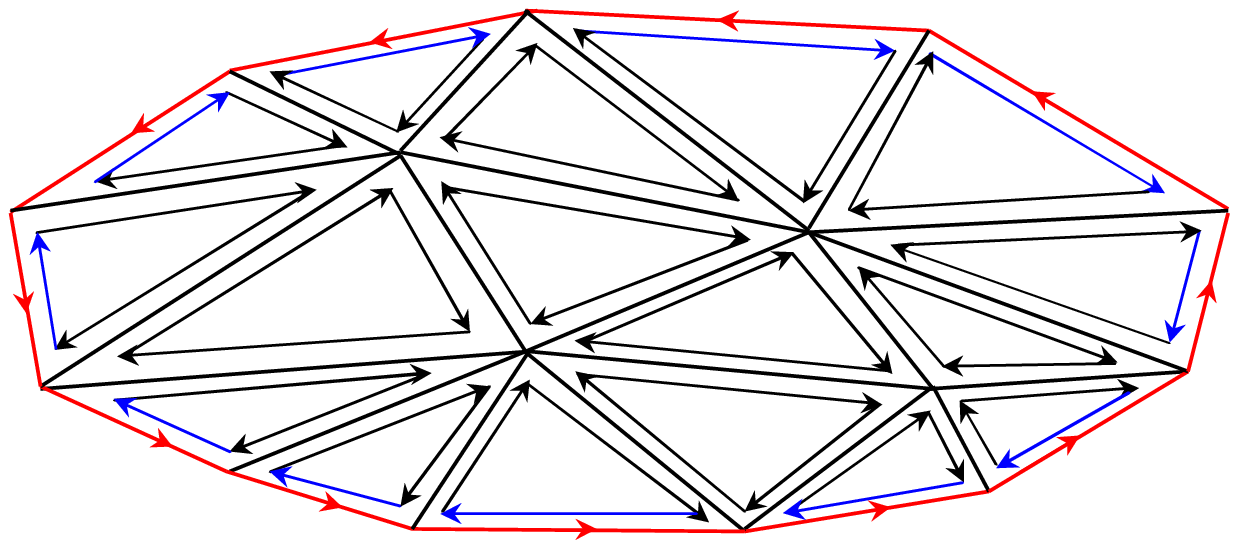}
\end{center}

\noindent{\small Figure 2.
Gravitational analog of the Wilson loop.
A vector is parallel-transported along the larger outer loop.
The enclosed minimal surface is tiled with parallel
transport polygons, here chosen to be triangles for illustrative
purposes.
For each link of the dual lattice, the elementary parallel transport
matrices $R (s,s')$ are represented by arrows. 
In spite of the fact that the (Lorentz) matrices ${\bf R}$ can fluctuate 
strongly in accordance with the local geometry, two contiguous, oppositely
oriented arrows always give $ R R^{-1} = 1$.
\medskip}

If that is true, then one can define an appropriate coordinate scalar 
by contracting the above rotation matrix ${\bf R}(C)$ 
with the some appropriate bivector, namely
\beq
W ( C ) \; = \; \omega_{\alpha\beta}(C) \, R^{\alpha\beta} (C) 
\label{eq:latt-wloop1}
\eeq
where the bivector,
$\omega_{\alpha\beta} (C )$, is intended as being representative of the overall
geometric features of the loop.

In the quantum theory one is of course interested in the average
of the above loop operator $W (C)$, as in
Eq.~(\ref{eq:wloop_sun}).
The previous construction is indeed quite analogous to the
Wilson loop definition in ordinary lattice gauge theories
[2], where it is defined
via the trace of path ordered products of $SU(N)$ color rotation matrices.
In gravity though the Wilson loop does not give any information
about the static potential [15,16].
It seems that the Wilson loop in gravity provides instead
some insight
into the large-scale curvature of the manifold, just
as the infinitesimal loop contribution entering the lattice action
of Eqs.~(\ref{eq:ilatt}) and (\ref{eq:regge-compact})
provides, through its averages, insight into the very short distance,
local curvature.

Of course for any continuum manifold one can define locally
the parallel transport of a vector around a near-planar loop
$C$.
Indeed parallel transporting a vector around a closed loop represents
a suitable operational way of detecting curvature locally.
If the curvature of the manifold is small, one can treat the larger loop
the same way as the small one; then the expression of Eq.~(\ref{eq:latt-wloop})
for the rotation matrix ${\bf R} (C )$ associated with a near-planar
loop can be re-written in terms of a surface
integral of the large-scale Riemann tensor, projected along the surface
area element bivector $A^{\alpha\beta} (C )$ associated with the loop,
\beq
R^{\mu}_{\;\; \nu} (C) \; \approx \; 
\Bigl [ \, e^{\half \int_S 
R^{\, \cdot}_{\;\; \cdot \, \alpha\beta} \, A^{\alpha\beta} ( C )} 
\Bigr ]^{\mu}_{\;\; \nu}  \;\; .
\eeq
Thus a direct calculation of the Wilson loop provides a way of determining
the {\it effective} curvature at large distance scales, even in the case
where short distance fluctuations in the metric may be significant.
Conversely, the rotation matrix appearing in the elementary Wilson loop of 
Eqs.~(\ref{eq:loop-rot1}) and (\ref{eq:rot-hinge}) only provides 
information about the parallel transport of vectors around
{\it infinitesimal} loops, with size comparable to the ultraviolet cutoff.

Let us now look in detail at how to construct a Wilson loop in quantum gravity. 
Since this involves finding the expectation value of a product of rotation 
matrices round a loop, the natural procedure is to treat these rotation 
matrices as variables and to integrate over their product, weighted by the 
exponential of minus the Regge action. The expression for this action has been given 
in terms of functions of the edge lengths, but an alternative [11,12,17]  is 
to find an expression for it in terms of the rotation matrices. For the dual 
loop around each hinge, the product of the rotation matrices gives the 
exponential of the deficit angle, $\delta$, times the rotation generator, $U$, 
(see Eq.~(\ref{eq:rot-hinge})) and we need to find a way of extracting the deficit 
angle from this product of 
matrices, at the same time as constructing a scalar function to be averaged. 
The obvious way of doing this is to contract the product of the $R$-matrices 
with the rotation generator, $U$, and then take the trace. This is equivalent 
to the action used in [11,12,17] (see also [18,19] for a hypercubic lattice
formulation), obtained
by contracting the elementary rotation matrix ${\bf R}(C)$ 
of Eq.~(\ref{eq:rot-hinge}), 
with the hinge bivector of Eq.~(\ref{eq:bivector-d}),
\beq
I_{\rm com} (l^2) \; = \; - \; {\frac {k} {2}} \, \sum_{\rm hinges \; h} \, A_h \, 
U_{\alpha\beta}(h) \, R^{\alpha\beta} (h) 
\label{eq:regge-compact}
\eeq
The above construction can be regarded as analogous to Wilson's lattice
gauge theory, for which the action also involves traces of products of
$SU(N)$ color rotation matrices [2]. This contraction produces the 
sine of the deficit angle times the area of the triangular hinge and so
for small deficit angles it is equivalent to the Regge action.
However, in general, away from a situation of small curvatures,
the two lattice action are not equivalent,
as can be seen already in two dimensions.

At this stage, we choose to differ from the choice made in [17],
for the simple reason that when we come to evaluate Wilson loops, the final result 
often involves the trace of the bivector $U$, which is zero.
Therefore, instead of 
contracting with $U$, we use a linear combination of it and the unit matrix. 
In particular, we take the contribution to the action, of a hinge, labelled $h$, 
to be

\beq
I_h \; = \; {\frac {k} {4}} \; A_h \; 
Tr [ (U_h \; + \; \epsilon \; I_4) \; ({\bf R}_h \; - \; {\bf R}^{-1}_h) ],  
\label{eq:ihinge}
\eeq

\noindent 
where $\epsilon$ is an arbitrary multiple of the unit matrix in four dimensions. 
We have subtracted the inverse of the rotation matrix for the hinge for reasons 
that will become apparent when we evaluate Wilson loops, and it also plays an 
important role in the action. At the end, we shall be interested in the limit 
of small but non-zero $\epsilon$.

The classical action may be evaluated as follows. Since $\bf{R}$ equals the exponential of 
$\delta$ times $U$, it may be expanded in a power series in $\delta$, which is 
then contracted with the $(U + \epsilon I_4)$ and the trace taken. 
We use 

\beq
 Tr(U^{2n+1}) \; = \; 0 \; ; \;\;\;\;  Tr(U^{2n}) \; = \; 2 \; (-1)^n,
\eeq
\noindent
to show that 

\beq
I_h \; = \; - \; k \; A_h \; \sin(\delta_h),
\label{eq:sin-ac}
\eeq 

\noindent 
independently of the value of the parameter $\epsilon$. (The unit matrix times
the even terms in the power series expansion produces a 
$ \, cos(\delta) \, $ term, but this 
cancels between the $\bf{R}$ and the $\bf{R}^{-1}$ contributions.) 
Thus $\epsilon$ is in fact an arbitrary parameter, 
which can be conveniently taken to be non-zero, as we shall see.

There is now a slight amount of freedom in how we define the Wilson loop, for a path 
$C$ in the dual lattice of a simplicial space. The main choices seem to be

\beq
(i) \;\;\;  W(C) \; = \; < \; Tr(R_1 \; R_2 \; ... \; R_n) \; >;
\eeq


\beq
(ii) \;\;\;  W(C)  \; = \; < \; Tr[(U_C \; + \; \epsilon \; I_4) \; 
R_1 \; R_2 \; ... \; ... \; R_n] \; > \; .
\eeq

\noindent Here the $R_i$ are the rotation matrices along the path; in (ii), there is a 
factor of $(U_C + \epsilon I_4)$, containing some \lq\lq average" direction bivector, 
$U_C$, for the loop, which, after all, is assumed to be almost planar.
The position of the $U_C$ term in the product of $R_i$'s is not arbitrary; 
to give a unique answer, it needs to be placed before an $R$ which begins
one of the plaquette contributions to the action.

We would like to take as independent fluctuating variables the rotation matrices 
$R_i$ and the loop bivectors $U_i$, in a first order formalism similar in
spirit to that used in [17].
This last statement clearly requires some clarification, as both the rotation matrices
and the loop bivectors depend on the choice of the original edge lengths,
as well as on the orientation of the local coordinate system, and
cannot therefore in general be considered as independent variables
(as should have already been clear from the detailed discussion of the properties
of rotation matrices given in the previous section).
On the lattice strong edge length fluctuations get reflected in large fluctuations
in the local geometry, which in turn imply large
correlated fluctuations in both the deficit angles and in the
orientations of the elementary loop.
It would therefore seem at first that one would have to integrate over
both sets of coupled variables simultaneously, with some non-trivial measure derived from
the original lattice measure over edge lengths, which in turn would make the problem
of computing the Wilson loop close to intractable, even in the strong coupling limit.
In particular one has to take notice of the fact that the lattice deficit angles and
the loop bivectors are related to the metric and connection, as they
appear in a first order formulation, in a rather non-trivial way.

But there are two important aspect that come into play when evaluating
the expectation value of the gravitational Wilson loop for strongly coupled
gravity, the first
one being that the overall geometric features of the large near-planar loop provide
a natural orientation, specified for example by a global loop bivector $U_C$.
As will become clear from explicit calculations given below,
in the strong coupling limit the tiling of the large Wilson loop surface
by elementary parallel transport loops, which in general have
random orientations, {\it requires} that their
normals be preferentially oriented perpendicular to the plane of
the loop, since otherwise a non-minimal surface must result, which
leads to a necessarily higher order contribution in the strong coupling limit.

In the case of a hinge surrounded by the large loop with bivector $U_C$,
one is therefore allowed to write for the bivector operator $U_h$ associated with
that hinge, labelled by $h$,
\beq
U_h \; = \; U_C \, + \, \delta U_h
\eeq
where $\delta U_h $ is the quantum fluctuation associated with hinge bivector at $h$.
But assuming the fluctuation in $\delta U_h$ to be zero is an unnecessarily strong
requirement, and in the following it will be sufficient to take $< \delta U_h > =0 $
and $ < ( \delta U_h) ^2 > \neq 0$, which can be regarded as a mean-field
type treatment for the loop bivectors.
It will be important therefore in the following to keep in mind
this distinction between the fluctutating hinge bivector $U_h$, 
and its quantum average.

The second important aspect of the calculation is that at strong coupling the
edge lengths, and therefore the local geometry, fluctuate in a way that is
uncorrelated over distances greater than a few lattice spacing.
Thus, mainly due to the ultralocal nature of the gravitational
lattice measure at strong coupling, the fluctuations
in the $U's$ can be taken as essentially uncorrelated as well,
again over distances greater than a few lattice spacings, which further
simplifies the problem considerably.

One would expect that for a geometry fluctuating strongly
at short distances (corresponding therefore to the small $k$ limit) the
infinitesimal parallel transport matrices
$ R (s,s')$ should be distributed close to randomly, with a measure close
to the uniform Haar measure, and with little correlation between
neighboring hinges.
In such instance one would have for the local quantum averages of the
infinitesimal lattice parallel transports $< R > = 0$, but 
$<  R \;  R^{-1} > \neq 0$, which would require, for a non-vanishing
lowest order contribution to the Wilson loop, that the loop at least
be tiled by elementary loops with action contributions from Eqs.~(\ref{eq:ilatt}) 
or (\ref{eq:regge-compact}), thus forming a minimal surface spanning the loop.
Then, in close analogy to the Yang-Mills case of Eq.~(\ref{eq:wloop_sun1})
(as a general reference, see for example [20]),
the leading contribution to the gravitational Wilson loop
would be expected to follow an area law,
\beq
< W( C ) > \; \sim \; 
{\rm const.} \, k^{A (C)} \; \sim \; \exp ( - A(C) / \xi^2 ) 
\label{eq:wloop_curv}
\eeq
where $A(C))$ is the minimal physical area spanned by the near-planar loop $C$,
and $\xi$ the gravitational correlation length, equal to 
$\xi = 1 /\sqrt{ \vert \ln k \vert }$ for small $k$.
For a close-to-circular loop of perimeter $P$ one would use 
$A (C) \approx P^2 / 4 \pi$.

We choose now to focus on the Euclidean case in four dimensions, where the 
rotation matrices will be elements of $SO(4)$.
In evaluating the averages over the rotation matrices in the expectation values in the Wilson
loops, the integrations we have to perform will be of the form 

\beq
\int \, \left ( 
\prod_{i=1}^n d \mu_H (  R_i ) \right ) \, 
Tr [...(U_j \, + \, \epsilon \, I_4) ... R_k ...]
\, \exp \left (
- \, {\frac {k} {4}} \, \sum_{\rm hinges \, h} A_h \, 
Tr [(U_h \, + \, \epsilon \, I_4) \, ({\bf{R}}_h 
\, - \, {\bf{R}}^{-1}_h)
 \, ] \right ) \, / \,  {\cal N}, 
\eeq   

\noindent where the normalization factor is given by

\beq 
{\cal N} \, = \, 
\int \, \left ( \prod_{i=1}^n d \mu_H (  R_i ) \right ) \, 
\, 
\exp \left ( - \, {\frac {k} {4}} \, \sum_{\rm hinges \, h} A_h \, 
Tr [(U_h \, + \, \epsilon \, I_4) \, ({\bf{R}}_h \, - \, {\bf{R}}^{-1}_h) ]
\, \right )    
\eeq

\noindent This factor will be omitted from subsequent expressions, for 
notational simplicity.

For smooth enough geometries, with small curvatures, the 
rotation matrices can be chosen to be close to the identity.
Small fluctuations in the geometry will then imply small deviations
in the $ R$'s from the identity matrix.
However, for strong coupling ($k \rightarrow 0$)
the usual lattice measure $\int d \mu (l^2)$ [21] does not significantly 
restrict fluctuations in the lattice metric field.
As a result we will assume that these fields can be regarded, 
at least in this regime, as basically unconstrained random variables, only subject
to the relatively mild constraints implicit in the measure $d \mu (l^2) $.
Thus as $k \rightarrow 0$, the geometry is generally far from smooth,
since there is no coupling term to enforce long range
order (the coefficient of the lattice Einstein term goes to zero),
and one has as a consequence large local fluctuations in the geometry.
The matrices ${\bf R}$ will therefore fluctuate with the local
geometry, and average out to zero, or a value close to
zero, in the sense that, for example, the $SO(4)$ rotation
\beq
 R_\theta = \left ( \matrix{ 
\cos \theta & - \sin \theta & 0 & 0 \cr
\sin \theta &   \cos \theta & 0 & 0 \cr 
0 & 0 & 1 & 0 \cr     0 & 0 & 0 & 1 \cr 
} \right )
\eeq 
averages out to zero when integrated over $\theta$. 
In general an element of $SO(n)$ is described by $n(n-1)/2$ independent 
parameters, which in the case at hand can be conveniently chosen as the six
$SO(4)$ Euler angles.
The uniform (Haar) measure over the group is then
\beq
d \mu_H (R) = {1 \over 32 \pi^9 }
\int_0^{2 \pi} d \theta_1 \int_0^{\pi} d \theta_2 \int_0^{\pi} d \theta_3
\int_0^{\pi} d \theta_4 \sin \theta_4 \int_0^{\pi} d \theta_5 \sin \theta_5
\int_0^{\pi} d \theta_6 \sin^2 \theta_6
\eeq
This is just a special case of the general $n$ result, which reads
\beq
d \mu_H ( R ) = 
\left ( \prod_{i=1}^n \Gamma (i/2) / 2^n \, \pi^{n(n+1)/2} \right )
\prod_{i=1}^{n-1} \prod_{j=1}^i 
\sin^{j-1} \theta_{\; j}^i \, d \theta_{\; j}^i
\eeq
with $ 0 \le \theta_{\; k}^1 < 2 \pi $, $ 0 \le \theta_{\; k}^j < \pi $ [22].

These averaging properties of rotations are quite similar to what 
happens in $SU(N)$ Yang-Mills theories,
or even more simply in (compact) QED, where the analogs of the $SO(d)$ rotation
matrices ${\bf R}$ are phase factors $U_{\mu}(x)=e^{iaA_{\mu}(x)}$.
There one has the trivial group averaging property 
$\int { d A_{\mu} \over 2 \, \pi } \, U_{\mu}(x) = 0 $
and 
$\int { d A_{\mu} \over 2 \, \pi } \, U_{\mu}(x) \, U^{\dagger}_{\mu}(x) = 1 $.
In addition, for two contiguous closed paths $C_1$ and $C_2$
sharing a common side one has
\beq
e^{ i \oint_{C_1} {\bf A \cdot dl } } \, e^{ i \oint_{C_2} {\bf A \cdot dl } }
\; = \; e^{ i \oint_{C} {\bf A \cdot dl } }
\; = \; e^{ i \int_S {\bf B \cdot n } \, dA } \;\; ,
\eeq
with $C$ the slightly larger path encircling the two loops.
For a closed surface tiled with many contiguous infinitesimal closed loops
the last expression evaluates to $1$, due to the
divergence theorem. 
In the lattice gravity case the discrete analog of this last result
represents the (exact) lattice
analog of the contracted Bianchi identities [23].

In practice, luckily, we do not have to explicitly integrate over 
$SO(4)$ angles, but rather use 
the following properties of the Haar measure, normalized to one, on the group:

\beq
\int d_H R \; = \; 1 \; ;
\eeq

\beq
\int d_H R \; \; Tr(A \; R) \; Tr(R^{-1} \; B) \; = \; 
\frac{1}{4} \; Tr(A \; B) \; ,
\eeq

\noindent for arbitrary $4 \times 4$ matrices $A$ and $B$, which also implies

\beq
\int d_H R \; \; R_{ij} \; R^{-1}_{kl} \; = \; 
\frac{1}{4} \; \delta_{il} \; \delta_{jk} \; .
\eeq

As stated previously, we will regard the individual hinge bivectors $U_h$ as
aligned on average with the Wilson loop bivector $U_C$.
Alternatively, one can perform the following simple exercise, in which
one assumes each hinge bivector $U_h$ is aligned in general
(arbitrary) directions, and performs an integration over those directions.
Here we will given an example of such a calculation.

Of course the normals to a 2-d loop in a 4-d space form a plane, rather than
a single direction, so in general what one needs to do is an integration
over the directions in that plane. However in this example, we are not
restricting the plane of the loop, so the integration is a
four-dimensional one over all possible directions. It is performed as follows: 
in the definition of $U_{\alpha \beta}$ in Eq.~(\ref{eq:bivector}), we put

\beq
l_1^{\alpha} = |l_1| a^{\alpha}, \;\;\;\;  l_2^{\beta} =  |l_2| b^{\beta},
\eeq

\noindent with $a_{\gamma} a^{\gamma} = b_{\gamma} b^{\gamma} = 1$. 
Since $A_h = {\frac {1} {2}} |l_1| |l_2| \sin \phi$, where $\phi$ is 
the angle between the vectors $l_1$ and $l_2$, the dependence on the
magnitudes of those two vectors cancels from the expression for $U$ 
and we obtain

\beq
U_{\alpha \beta} = 
{\frac {\epsilon_{\alpha \beta \gamma \delta} \, a^{\gamma} b^{\delta}} {\sin \phi}},
\eeq

\noindent with $\cos \phi = a_{\gamma} b^{\gamma}$. 
The integration of $U_{\alpha \beta}$ then becomes

\beq
{\frac {4} {\pi^4}} \, {\prod_{i=1}^4 \int_{-1}^1 da_i} \, 
{\prod_{j=1}^4 \int_{-1}^1 db_j} 
\frac 
{\epsilon_{\alpha\beta\gamma\delta} \, a^{\gamma} b^{\delta} \, 
\delta ( a_{\lambda} a^{\lambda} - 1 ) \, \delta ( b_{\rho} b^{\rho} - 1 ) } 
{\sqrt{1 - (a_\mu b^\mu)^2} },
\eeq

\noindent where the coefficient in front of the integral ensures that
the measure is normalized to 1.

\vskip 40pt

\section{Evaluation of Wilson loops and behavior for large loops}

\vskip 20pt

In this section, we first evaluate $<W>$ for some simple loops and then 
discuss the general behavior for arbitrary loops, ending with a 
consideration of the asymptotic behavior for large loops. We shall work 
out  $<W>$ for the two possible definitions listed in the previous 
section.

\vskip 40pt

\subsection{Loop round a single hinge}

\vskip 20pt

Consider a single hinge of area $A$, at which four 4-simplices meet
(see Figure 3.).
The loop $C$ will consist of four segments between the Voronoi centers of the 
simplices. Let the rotation matrices on these segments be 
$R_1, R_2, R_3, R_4$, and the rotation generator for the hinge $U$. Then 
we have either

\beq
(i) \;\;\; W(C) \; = \; < \; Tr(R_1 \; R_2 \; R_3 \; R_4) \; >
\eeq
or
\beq
(ii) \;\;\; W(C) \; = \; < \; Tr [(U \; + \; \epsilon \; I_4) \; 
R_1 \; R_2 \; R_3 \; R_4] \; >.
\eeq

\begin{center}
\epsfxsize=4cm
\epsfbox{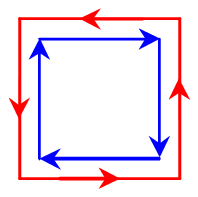}
\end{center}

\noindent{\small Figure 3.
A parallel transport loop with four oriented links on the boundary.
The parallel transport matrices ${\bf R}$ along the links, 
represented here by arrows, appear in pairs and are 
sequentially integrated over using the uniform measure.
\medskip}

Since the only non-vanishing contribution to the integration over the $R$'s will come 
from the product of an $R_i$ with the corresponding $R^{-1}_i$, then the 
lowest order contribution in $k$ will come from the term in the expansion 
of the exponential of minus the action which is linear in ${\bf R}^{-1}$. 
Thus in case (i) we obtain

\beq
{\frac {k} {4}} \; A \; \int \; d_HR_1 \; d_HR_2 \; d_HR_3 \; d_HR_4 \; 
Tr(R_1 \; R_2 \; R_3 \; R_4) \;   
Tr [(U \; + \; \epsilon \; I_4) \; R^{-1}_4 \; R^{-1}_3 \; R^{-1}_2 \; R^{-1}_1 ],
\eeq

\noindent and a similar expression for (ii), with the extra factor of 
$(U \; + \; \epsilon \; I_4)$ inserted.
Integration over the $R$'s results in

\beq
(i) \;\;\; {\frac {k} {4}} \; {\frac {1} {4}} \; 
A \; Tr(U \; + \; \epsilon \; I_4) \; = \; {\frac {k} {4}} \; A \; \epsilon
\eeq 

or

\beq
(ii) \;\;\; {\frac {k} {4}} \; {\frac {1} {4}} \; 
A \; Tr(U^2 \; + \; \epsilon^2 \; I_4) \; 
= \; - \; {\frac {k} {8}} \; A \; (1 \; - \; 2 \; \epsilon^2) \; ,
\eeq

\noindent and integration over the $U$'s (a sum over all possible orientations
of the loop) is trivial.

\vskip 40pt

\subsection{Loop round two hinges}

\vskip 20pt

Suppose now that the loop goes around two adjacent hinges, with rotation 
bivectors $U_1, U_2$. The $R$ matrices going clockwise round the loop 
in Figure 4., 
starting at the top left hand corner, are labelled $1,2,5,6,7,4$ and the one 
between the two loops is $R_3$.

\begin{center}
\epsfxsize=6cm
\epsfbox{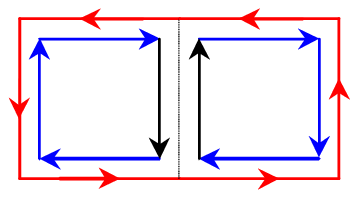}
\end{center}

\noindent{\small Figure 4.
A parallel transport loop with six oriented links on the boundary.
\medskip}

Then the possible values of the Wilson loop are 

\beq
(i) \;\;\; W(C) \; = \; < \; Tr(R_1 \; R_2 \; R_5 \; R_6 \; R_7 \; R_4) \; >,
\eeq


or

\beq
(ii) \;\;\; W(C) \; = \; 
< \; Tr[(U_C \; + \; \epsilon \; I_4) \; 
R_1 \; R_2 \; R_5 \; R_6 \; R_7 \; R_4] \; >.
\eeq

This time, the lowest order non-zero contribution will come from 
the term in the expansion of the exponential which is quadratic  
and involves the product of the ${\bf R}^{-1}$'s from the two hinges, 
so for (i) the integration over the $R$'s gives

\bea
&& {\frac {k^2} {16}} \, A_1 \, A_2 \, \int \, 
d_HR_1 \, d_HR_2 \, d_HR_3 \, d_HR_4 \,  d_HR_5 \, d_HR_6 \, d_HR_7 \; 
Tr(R_1 \, R_2  \, R_5 \, R_6 \, R_7 \, R_4) \, 
\nonumber \\
&& \;\;\; \times \,
Tr [(U_1 \, + \, \epsilon \, I_4) \, 
R^{-1}_4 \, R^{-1}_3 \, R^{-1}_2 \, R^{-1}_1] \,
Tr [(U_2 \, + \, \epsilon \, I_4) \, 
R_3 \, R^{-1}_7 \, R^{-1}_6 \, R^{-1}_5 ] \; , 
\eea
and a similar expression for (ii). 
Evaluation of the integrals leads to
\beq
(i) \;\;\; {\frac {k^2} {16}} \; {\frac {1} {4^3}} \; A_1 \; A_2  \; 
Tr (U_1  \; + \; \epsilon \; I_4) \; Tr(U_2 \; + \; \epsilon \; I_4) 
\; = \; {\frac {k^2} {64}} \; A_1 \; A_2 \; \epsilon^2 \; ,
\label{eq:tr-i}
\eeq 


or

\beq
(ii) \;\;\; {\frac {k^2} {16}} \; {\frac {1} {4^3}} \; A_1 \; A_2  \; 
Tr [(U_C \; + \; \epsilon \; I_4) \; 
(U_1  \; + \; \epsilon \; I_4) ] \; Tr (U_2 \; + \; \epsilon \; I_4) 
\; = \; 
{\frac {k^2} {256}} \; A_1 \; A_2 \; \epsilon \; 
[Tr(U_C \; U_1) \; + \; 4 \; \epsilon^2 ] \; .
\label{eq:tr-ii}
\eeq 

The integration over the $U$'s, and therefore over the loop's
orientation, in Eq.~(\ref{eq:tr-i}) is trivial. 
In Eq.~(\ref{eq:tr-ii}), we either set $U_1 = U_C + \delta U_1$, with
$<\delta U_1>=0$ after which the sum over the loop's orientation also 
becomes trivial since $ Tr(U_C U_1) =  -2$, 
or we can use the following technique to evaluate
the integral of $Tr(U_C U_1)$: we assume (as usual) that $U_C$ corresponds
to a planar or almost-planar loop and we choose Cartesian coordinates 
(in the tangent space to it) such that the loop is in the $(1,2)$ plane.
We take the generator of $U_C$ to be a right-angled triangle in 
the $(3,4)$ plane, with $l{_C}{_1} = (0,0,1,0)$, $l{_C}{_2} = (0,0,0,1)$
and $A_C = 1/2$. 
It can then be shown that

\beq
Tr(U_C U_1) = - {\frac {1} {A}} 
\left [ (l_1)_3 (l_2)_4 - (l_1)_4 (l_2)_3 \right ] \; ,
\eeq

\noindent where $l_1$ and $l_2$ are edge-vectors for the triangular
hinge for $U_1$ and $A$ is the triangle's area. 
The integral to be evaluated here is therefore

\beq
- \frac {8} {\pi^4} \, 
\prod_{i=1}^4 \int_{-1}^1 da_i \, 
\prod_{j=1}^4 \int_{-1}^1 db_j 
\frac { (a_3 b_4 - a_4 b_3) \, 
\delta( a_\lambda a^\lambda - 1 ) \, \delta( b_{\rho} b^{\rho} - 1 ) }
{\sqrt{1 - (a_\mu b^\mu)^2} }.
\eeq

\noindent It can be shown that this integral is zero 
(an indication that this might be so is the antisymmetry in $a$ and $b$ 
in the integrand), and so Eq.~(\ref{eq:tr-ii}) becomes identical to 
Eq.~(\ref{eq:tr-i}), apart from an extra power of $\epsilon$.

\vskip 40pt

\subsection{Loop with one internal hinge}

\vskip 20pt

We now need to consider the situation where the Wilson loop goes around a number 
of hinges and there is at least one internal hinge, i.e. a hinge where the elementary 
loop surrounding it is not part of the Wilson loop. For simplicity, we shall consider 
the case of one such loop. For the labelling of the rotation matrices and the hinges, 
the reader can annotate Figure 5. in a way consistent with the expressions below.

\begin{center}
\epsfxsize=8cm
\epsfbox{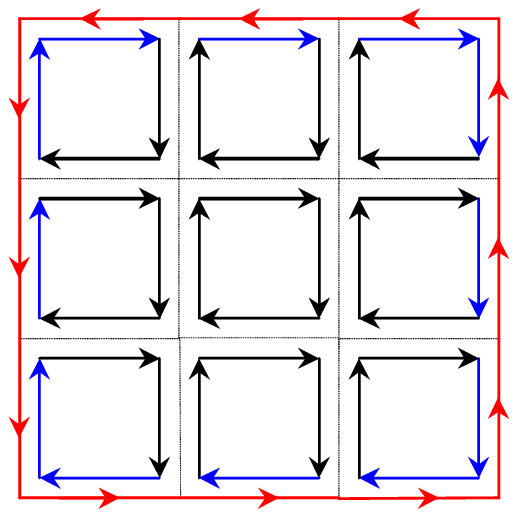}
\end{center}

\noindent{\small Figure 5.
A larger parallel transport loop with twelve oriented links on the boundary.
As before, the parallel transport matrices along the links appear
in pairs and are sequentially integrated over using the uniform
measure. The new ingredient in this configuration is an elementary
loop at the center not touching the boundary.
\medskip}

In this case, the lowest order contribution comes from a ninth-order 
term in the expansion of the exponential of the action. We obtain the following 
results in the two cases:

\beq
(i) \;\;\; {\frac {k^9} {4^9}} \; {\frac {1} {4^{17}}} \; 
\left (\prod_{i=1}^9 \; A_i \right ) \;  
\left (\prod_{i=1}^9 \; Tr(U_i \; +
\; \epsilon \; I_4) \right ) \; = \; 
{\frac {k^9} {4^{17}}} \; \left (\prod_{i=1}^9 \; A_i \right ) \; \epsilon^9,
\label{eq:tr-ia}
\eeq


or  

\bea
(ii) && {\frac {k^9} {4^9}} \, {\frac {1} {4^{17}}} \,  
\left (\prod_{i=1}^9 \, A_i \right ) \, Tr [(U_C + \epsilon I_4) \, 
(U_1 + \epsilon I_4) ] \, \left (\prod_{i=2}^9 \, 
Tr(U_i\ + \epsilon I_4) \right ) \, 
\nonumber \\
&& \;\; = \, {\frac {k^9} {4^{18}}} \, \left (\prod_{i=1}^9 \, A_i \right ) 
\, \epsilon^8 \, [ Tr(U_C U_1) + 4 \epsilon^2 ].
\label{eq:tr-iia}
\eea

Integration over the U's is trivial for Eq.~(\ref{eq:tr-ia}).
In Eq.~(\ref{eq:tr-ii}), we either set $U_1 = U_C + \delta U_1$, with
$<\delta U_1>=0$ after which the sum over the loop's orientation also 
becomes trivial, or we can integrate over the relative
orientation of $U_1 $ relative to $U_C$ using the integral formula
given at the end of the previous section, after which
Eq.~(\ref{eq:tr-iia}) gives the same as Eq.~(\ref{eq:tr-ia}) 
but for an extra power of $\epsilon$.
The above result also shows that it is better to take $\epsilon >0 $,
otherwise the answer vanishes to this order.
But this is not a problem, as the correct lattice action is recovered
irrespective of the value of $\epsilon$, as shown earlier in
Eq.~(\ref{eq:sin-ac}).

\vskip 40pt

\subsection{Large loop}

\vskip 20pt

The value of a Wilson loop, in the case when the loop is very large and surrounds 
$n$ hinges, can be seen to be roughly of the general form

\beq
 {\frac {k^n} {4^{2n}}} \, 
\left (\prod_{i=1}^n \, A_i \right ) \, \epsilon^{\alpha} \, 
[ p \, + \, q \, \epsilon^2 ]^{\beta},
\eeq

\noindent where $\alpha + \beta = n$. If ${\bar A}$ is of the order of the 
geometric or arithmetic mean of the individual loops, this can be approximated 
by

\beq
\left ({\frac {{k \,  {\bar A}}} {16}} \right )^n \, \epsilon^{\alpha} 
\, [ p \, + \, q \, \epsilon^2 ]^{\beta}.
\eeq

\noindent
The above result shows again that one should take $\epsilon >0 $,
otherwise the answer vanishes to this order.
As mentioned previously this is quite legitimate, 
as the correct lattice action is recovered
irrespective of the value of $\epsilon$, as in Eq.~(\ref{eq:sin-ac}).
Then using $n = A_C /{\bar A}$, 
we may write the area-dependent first factor as

\beq
\exp [ \,( A_C / {\bar A}) \, \log (k \, {\bar A} / {16}) \, ] 
\, = \, \exp ( - \, A_C / {\xi }^2 )
\label{eq:expdecay}
\eeq

\noindent where 
$\xi = \sqrt{[{\bar A} / \vert \log (k \, {\bar A} / {16}) \vert]}$. 
Recall that this is
in the case of strong coupling, when $k \rightarrow 0$.
The above is the main result of this paper.
The rapid decay of the quantum gravitational Wilson loop as a function of the
area is seen here simply as a general and direct consequence of the
assumed disorder in the uncorrelated fluctuations of the parallel transport
matrices ${\bf R}(s,s')$ at strong coupling.

We note here as well that the correlation length $\xi$ is defined independently
of the expectation value of the Wilson loop.
Indeed a key quantity in gauge theories as well as gravity is the
{\it correlation} between different plaquettes, which in simplicial 
gravity is given by
\beq
< ( \delta \, A )_{h} \, ( \delta \, A)_{h'} > \; = \; {\displaystyle
\int d \mu (l^2) \,
( \delta \, A)_{h} \, ( \delta \, A)_{h'} \,
e^{k \sum_h \delta_h \, A_h }
\over
\displaystyle \int d \mu (l^2) \, e^{k \sum_h \delta_h \, A_h } } \;\; .
\label{eq:corr}
\eeq
In order to achieve
a non-vanishing correlation one needs, at least to lowest order,
to connect the two hinges by a narrow tube [24], so that
\beq
< ( \delta \, A )_{h} \, ( \delta \, A)_{h'} >_C \; \sim \;
\left ( k^{n_t} \right )^l \sim \; e^{- d(h,h') / \xi } \;\; ,
\label{eq:kxi}
\eeq
where $n_t \, l$ represents the minimal number of dual
lattice polygons needed to form a closed surface connecting the hinges $h$ and $h'$ 
(as an example, for a narrow tube
made out of cubes connecting two squares one has $n_t$=4).
In the above expression $ d(h,h')$ represents
the actual physical distance between the two hinges,
and the correlation length is given in this limit
($k \rightarrow 0 $) by
$ \xi \sim l_0 / n_t | \log k | $.
where $l_0$ is the average lattice spacing.
Here we have used the usual definition of the correlation length $\xi$,
namely that a generic correlation function
is expected to decay as $ \exp (- {\rm distance} / \xi) $ for
large separations. Figure 6. provides an illustration of the situation.

\begin{center}
\epsfxsize=10cm
\epsfbox{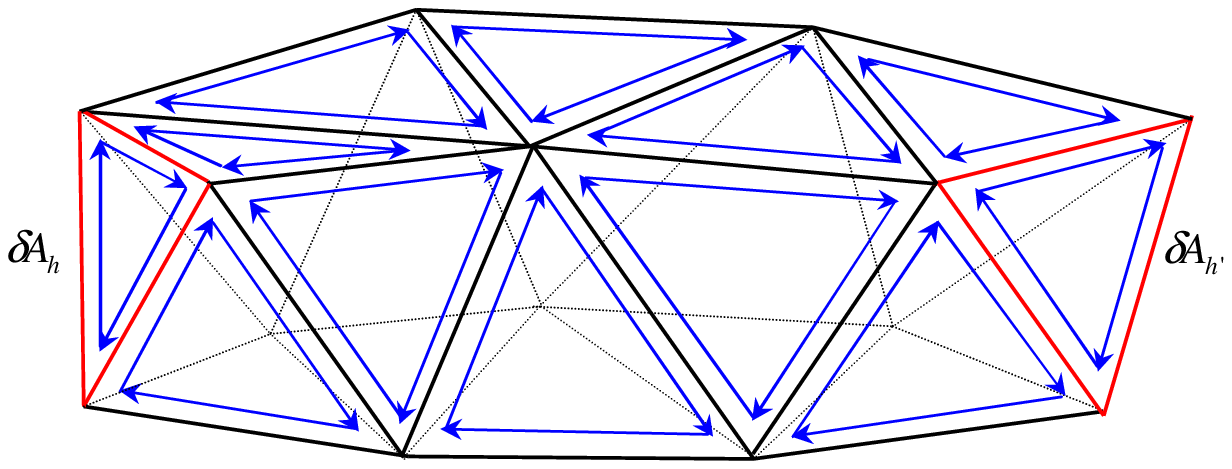}
\end{center}

\noindent{\small Figure 6.
Correlations between action contributions on hinge $h$
and hinge $h'$ arise to lowest order in the strong coupling expansions
from diagrams describing a narrow tube connecting the two hinges.
Here vertices represent points in the dual lattice, with
the tube-like closed surface tiled with parallel transport polygons.
For each link of the dual lattice, the $SO(4)$ parallel transport matrices
${\bf R}$ are represented by an arrow.
\medskip}

The strong coupling area law behavior predicted for a large Wilson loop in 
Eq.~(\ref{eq:expdecay}) should be compared with the results for this in 
numerical simulations of lattice gravity. For small deficit angles (small 
curvature), the action used in this paper [involving Eq.~(\ref{eq:ihinge})] 
is sufficiently close to the usual Regge action of Eq.~(\ref{eq:ilatt}) that 
the standard simulations can be used for comparison.
Universality arguments would suggest a similar behavior for the
gravitational Wilson loop for a wide class of lattice actions,
constructed so as to reproduce the Einstein-Hilbert continuum
action in the continuum limit.

\vskip 40pt

\section{Conclusions}

\vskip 20pt

In this paper, we have provided two possible constructions of Wilson loops 
in the gravitational case, and have given explicit calculations for small loops,
as well as deriving the asymptotic behavior for large loops.
The final step, which we will give here,
will be an attempt at providing an interpretation of this last and main
result in semiclassical terms.
As discussed in the introduction, the rotation matrix appearing in the gravitational
Wilson loop can be related classically to a well-defined physical process:
a vector is parallel transported around a large loop, and at the end it
is compared to its original orientation.
The vector's rotation is then directly related to some sort
of average curvature enclosed by the loop.
The total rotation matrix ${\bf R}(C)$ is given in general by a
path-ordered (${\cal P}$) exponential of the integral of the
affine connection $ \Gamma^{\lambda}_{\mu \nu}$ via
\beq
R^\alpha_{\;\; \beta} (C) \; = \; \Bigl [ \; {\cal P} \, \exp
\left \{ \oint_
{{\bf path \; C}}
\Gamma^{\cdot}_{\lambda \, \cdot} d x^\lambda
\right \}
\, \Bigr ]^\alpha_{\;\; \beta}  \;\; .
\label{eq:rot-cont}
\eeq
In such a semiclassical description of the parallel transport
process of a vector around a very large loop, one can re-express
the connection in terms of a suitable coarse-grained, or semi-classical, 
Riemann tensor, using Stokes' theorem
\beq
R^\alpha_{\;\; \beta} (C) \; \sim \;
\Bigl [ \;
\exp \, \left \{ \half \,
\int_{S(C)}\, R^{\, \cdot}_{\;\; \cdot \, \mu\nu} \, A^{\mu\nu}_{C} \;
\right \}
\, \Bigr ]^\alpha_{\;\; \beta}  \;\; ,
\label{eq:rot-cont1}
\eeq
where here $ A^{\mu\nu}_{C}$ is the usual area bivector associated with the
loop in question,
\beq
A^{\mu\nu}_{C} = \half \oint dx^\mu \, x^\nu .
\eeq
The use of semi-classical arguments in relating the above rotation matrix
${\bf R}(C)$
to the surface integral of the Riemann tensor assumes (as usual in the
classical context) that the curvature is slowly varying on the scale
of the very large loop.
Since the rotation is small for weak curvatures, one can write
\beq
R^\alpha_{\;\; \beta} (C) \; \sim \;
\Bigl [ \, 1 \, + \, \half \,
\int_{S(C)}\, R^{\, \cdot}_{\;\; \cdot \, \mu\nu} \, A^{\mu\nu}_{C}
\, + \, \dots \, \Bigr ]^\alpha_{\;\; \beta}  \;\; .
\label{eq:rot-cont2}
\eeq
At this stage one is ready to compare the above expression to the
quantum result
of Eq.~(\ref{eq:expdecay}), and in particular one should relate
the coefficients of the area terms, which
leads to the identification of the magnitude of the large scale
semiclassical curvature with the genuinely quantum
quantity $ 1  / \xi^2 $. Since one expression [Eq.~(\ref{eq:rot-cont2})] is a 
matrix and the other [Eq.~(\ref{eq:expdecay})] is a scalar, 
we shall take the trace after first 
contracting the rotation matrix with $(U_C \, + \, \epsilon \, I_4)$, as in our second 
definition of the Wilson loop, giving

\beq
W(C) \, \sim \, \Tr \left ( (U_C \, + \, \epsilon \, I_4) \, \exp \, 
\left \{ \, \half \,
\int_{S(C)}\, R^{\, \cdot}_{\;\; \cdot \, \mu\nu} \, A^{\mu\nu}_{C} \; 
\right \} \right ) .
\label{eq:wloop_curv1}
\eeq
           
\noindent  As is standard in simplicial gravity, we use 

\beq 
 R^{\, \alpha}_{\;\; \beta \, \mu\nu} \; = \; 
{\bar R} \; U^{\, \alpha}_{\;\; \beta} \; U_{\mu\nu} ,
\eeq

\noindent where ${\bar R}$ is some average curvature over the loop, and the 
$U$'s here will be taken to coincide with $U_C$. The trace of the product of 
$(U_C \, + \, \epsilon \, I_4)$ with this expression gives

\beq
Tr( {\bar R} \; U_C^2 \; A_C ) \; = \; - \; 2 \; {\bar R} \; A_C ,
\eeq

\noindent where we have used $ U_{\mu\nu} \, A^{\mu\nu}_{C} \, = 2 \, A_C$
(we choose the directions of the bivectors such that the latter
is true for all loops).
This is 
to be compared with the linear term from the other exponential expression, 
$- \, A_C / \xi^2 $. Thus the average curvature is computed to be of the 
order ${\bar R} \sim 1 / \xi^2 $, at least in the small $k$ limit [25].
An equivalent way of phrasing the last result is that $1 / \xi^2$ should
be identified, up to a constant of proportionality, with the
scaled cosmological constant $\lambda$, which can be regarded as a measure of
the intrinsic curvature of the vacuum.

One important aspect of lattice gravity, and of the estimate
for the large scale behavior of the Wilson loop given here, is that
it can be tested by numerical methods.
That is, the large scale behavior of the Wilson loop can in principle
be computed directly by evaluating the lattice path integral using 
numerical methods, thus bypassing entirely the need for a separate
treatment of the quantum fluctuations in the rotation matrices ${\bf R}$
and in the elementary loop bivectors $U$, as was done here in order
to obtain an analytical result in the strong coupling limit.
Furthermore in a numerical treatment of the Wilson loop one
is no longer restricted necessarily to the strong coupling limit.

We see that a  direct calculation of the Wilson loop for gravity can
provide an insight into whether the manifold is De Sitter or
anti-De Sitter {\it at large distances}
\footnote{In contrast, numerical studies of full lattice quantum gravity 
show that the {\it elementary} Wilson loop, describing the parallel
transport around a {\it single} hinge, provides evidence
for negative curvature at distance scales comparable to the 
ultraviolet cutoff [5].}.
Note that 
the definition of the gravitational Wilson loop is based on a surface
with a given boundary $C$,
in the simplest case the minimal surface spanning the loop.
It is possible though to consider other surfaces built out of elementary
parallel transport loops.
These will be considered elsewhere.

\vspace{40pt}

{\bf Acknowledgements}

The authors wish to thank Hermann Nicolai and the
Max Planck Institut f\" ur Gravitationsphysik (Albert-Einstein-Institut)
in Potsdam for very warm hospitality. 
They also wish to thank the referee for very helpful comments.
The work of HWH was supported in part by the Max 
Planck Gesellschaft zur F\" orderung der Wissenschaften.
The work of RMW was supported in 
part by the UK Particle Physics and Astronomy Research Council.

\vfill

\newpage

\end{document}